\documentclass[11pt]{article}

\usepackage[top=50mm, bottom=50mm, left=50mm, right=50mm]{geometry}

\usepackage{amsmath}
\usepackage{amssymb}
\usepackage{amsfonts}
\usepackage{mathtools}
\usepackage{mathrsfs}
\usepackage[scr = dutchcal]{mathalpha}
\usepackage{graphicx}
\usepackage{graphics}
\usepackage{float}
\usepackage{color}
\usepackage{fullpage}
\usepackage[normalem]{ulem} 
\usepackage{makeidx}
\usepackage{xspace}
\usepackage{authblk}
\usepackage{datetime}
\usepackage{microtype}
\usepackage{xpatch}
\usepackage{xstring}
\usepackage{fancyhdr}
\usepackage{mathrsfs}
\usepackage{esint}
\usepackage{upgreek}
\usepackage{siunitx}
\usepackage{chemformula}
\usepackage{tabularx, multirow}
\usepackage{threeparttable}
\usepackage[]{booktabs} 
\usepackage{float}
\usepackage[section]{placeins}
\usepackage{graphicx,xcolor}
\usepackage[colorlinks=false]{hyperref}
\usepackage[noabbrev,sort&compress]{cleveref}
\usepackage[font = footnotesize, labelfont=bf]{caption}
\usepackage{setspace}
\usepackage{cite}
\usepackage[square,numbers,sort&compress]{natbib}

\fancypagestyle{fpg}{%
	\lfoot{\footnotesize Preprint}%
	\cfoot{\normalsize\thepage}
	\rfoot{\footnotesize \textbf{2023}\href{}{}}%
}
\newcommand{\doi}[1]{DOI: \href{http://dx.doi.org/#1}{\nolinkurl{#1}}}
\makeindex

\author[1,$\dagger$]{Amin Ebrahimi}
\author[2]{Mohammad Sattari}
\author[1]{Aravind Babu}
\author[1]{Arjun Sood}
\author[2]{Gert-Willem~R.B.E.~R{\"o}mer}
\author[1]{Marcel~J.M.~Hermans}

\affil[1]{\textit{Department of Materials Science and Engineering, Faculty of Mechanical, Maritime and Materials~Engineering, Delft~University~of~Technology, Mekelweg~2, 2628CD~Delft, The~Netherlands}}
\affil[2]{\textit{Chair of Laser Processing, Department of Mechanics of Solids, Surfaces and Systems (MS\textsuperscript{3}), Faculty of Engineering Technology, University~of~Twente, Drienerlolaan~5, 7522NB~Enschede, The~Netherlands}}
\affil[$\dagger$]{Corresponding author: A.Ebrahimi@tudelft.nl (A. Ebrahimi)}

\title{\LARGE\textbf{Revealing the~effects of laser beam shaping on melt~pool~behaviour in conduction-mode laser melting}}
\date{}

\begin{document}

\maketitle
\thispagestyle{fpg}

\begin{abstract}
	Laser beam shaping offers remarkable possibilities to control and optimise process stability and tailor material properties and structure in laser-based welding and additive manufacturing. However, little is known about the~influence of laser beam shaping on the~complex melt-pool behaviour, solidified melt-track bead profile and microstructural grain morphology in laser material processing. A~simulation-based approach is utilised in the~present work to study the~effects of laser beam intensity profile and angle of incidence on the~melt-pool behaviour in conduction-mode laser melting of stainless steel 316L plates. The~present high-fidelity physics-based computational model accounts for crucial physical phenomena in laser material processing such as complex laser-matter interaction, solidification and melting, heat and fluid flow dynamics, and free-surface oscillations. Experiments were carried out using different laser beam shapes and the~validity of the~numerical predictions is demonstrated. The~results indicate that for identical processing parameters, reshaping the~laser beam leads to notable changes in the~thermal and fluid flow fields in the~melt pool, affecting the~melt-track bead profile and solidification microstructure. The~columnar-to-equiaxed transition is discussed for different laser-intensity profiles.
\end{abstract}

\bigskip
\newpage

\onehalfspacing
\section{Introduction}
\label{sec:intro}

Laser beam melting offers unique controllability and high spatial and temporal precision in manufacturing high-integrity metallic products and has been utilised in many industries~\cite{Herzog_2016}. Melting and solidification in laser welding and additive manufacturing considerably affect local microstructural features (\textit{i.e.}~solidification morphology, grain size and crystallographic texture)~\cite{Khare_2007,Roehling_2017,Roehling_2020,Shi_2020} and the~macroscale structure and properties of the~material~\cite{Mei_2017,Gruenewald_2021,FathiHafshejani_2022}. Studies have demonstrated that internal flow behaviour in melt pools has a~significant influence on melting and subsequent solidification~\cite{Ebrahimi_2021,Ebrahimi_2021_b}, hence it needs to be controlled effectively to achieve the~desired properties~\cite{Cooke_2020}. However, controlling internal flow behaviour in laser welding and additive manufacturing is challenging mainly due to the~complexity of molten metal flow dynamics and high sensitivity to changes in process parameters~\cite{Ebrahimi_2020,Ebrahimi_2022,Ebrahimi_2023}. Therefore, understanding underlying physical phenomena in laser melting (\textit{i.e.}~laser-matter interaction, heat and molten metal flow, and solid-liquid phase transformation) is crucial for effective control of the~melt-pool behaviour.

Previous studies suggest laser intensity profile as a~critical factor influencing the~processing window, melt-pool shape and dimension, and product quality in laser welding and additive manufacturing~\mbox{\cite{Ayoola_2019,Tenbrock_2020,Tumkur_2021,Mi_2022,Pamarthi_2023}}. Changes in the~laser intensity profile in such processes affect heating and cooling cycles by influencing convection in the~melt pool~\cite{Cloots_2016} and hence can lead to changes in thermal gradients and solidification growth rates~\cite{Kubiak_2015,Collins_2016,Bremer_2023}. Advancements in optics and laser technology have opened up new possibilities for modulating spatial and temporal laser intensity profiles with high resolution, which is beneficial to enhancing process stability and control in laser material processing. Various techniques have been developed to tailor laser beam shape, which are reviewed comprehensively elsewhere~\cite{Dickey_2018,Salter_2019}. However, the~influence of tuning spatial laser intensity profile (commonly known as laser beam shaping) on internal flow behaviour has been overlooked and only sparsely investigated, particularly concerning conduction-mode laser melting, which serves as the foundation for numerous welding and additive manufacturing techniques. Moreover, the~design and optimisation of laser intensity profile usually relies on trial-and-error experiments that are generally expensive and time-consuming~\cite{Bi_2023,Sundqvist_2016}. This highlights the necessity of developing reliable predictive models that facilitate design-space exploration and process optimisation within a reasonable timeframe.

Studies on the~effect of laser beam shaping in fusion welding and additive manufacturing, to~a~great extent, are experimental and focus primarily on measuring melt-pool shape and the~resulting solidification texture and microstructural properties (see for instance~\cite{Khare_2007,Kell_2012,Funck_2014,Ayoola_2019,Mi_2022}). High-fidelity numerical simulations supplement experimental studies by providing a~better insight into the~complex heat and fluid flow in melt pools and associated melting and solidification, supporting the~development of process design and optimisation for laser-based manufacturing. \citet{Han_2004} developed a~three-dimensional computational model and studied the~effects of employing a~rectangular and three different axisymmetric laser intensity profiles on the~evolution of thermal and fluid flow fields in stationary laser melting of a~stainless steel alloy (AISI~304). In their model, \citet{Han_2004} neglected the~influence of surface-active elements on Marangoni flow pattern. \citet{Safdar_2007} conducted three-dimensional numerical simulations to study the~effects of different rectangular beam shapes on thermal and fluid flow fields in laser melting of a~mild steel alloy (EN-43A). They neglected the~influence of surface-active elements on the Marangoni flow pattern in their model and assumed that the~melt-pool surface is flat and does not deform during the~process. Numerical predictions of laser beam welding of aluminium-copper alloys reported by~\mbox{\citet{Rasch_2019}} showed that the~average surface temperature and resulting recoil pressure changes notably when using different beam shapes with comparable laser powers. \mbox{\citet{Roehling_2017,Roehling_2020}} studied the~effects of elliptical laser beam shapes on the~resulting microstructure of stainless steel 316L in laser powder-bed additive manufacturing and demonstrated experimentally that microstructure grain nucleation and morphology could be controlled using the~laser beam shaping technique. Later, \citet{Shi_2020} developed a~cellular automaton model to predict solidification grain structure using elliptical laser beam shapes in laser powder-bed additive manufacturing and employed a~thermal-fluid model to predict the~evolution of the~thermal field during the~process. The~influence of surface-active elements on Marangoni flow pattern was neglected in their thermal-fluid simulations. Focusing on conduction-mode laser melting of Ti6Al4V alloy, \mbox{\citet{Abadi_2021b}} numerically studied the~effects of using elliptical and circular beams with a~Gaussian intensity profile on molten metal flow behaviour and melt-pool shape. In these studies, the~effects of surface morphology, laser incidence angle and temperature on local energy absorption are neglected and the~value of absorptivity is assumed to be constant. Neglecting these factors limits the general applicability of the model in predicting melt-pool behaviour for different laser intensity profiles and various laser material processing scenarios. Despite the~great potential of laser beam shaping for improving process control and final product quality, very little is known about the~influence of various laser beam shapes on heat and fluid flow in melt pools and the~resulting bead quality.

The~present work focuses on studying laser beam melting of metallic substrates using different beam shapes, intensity profiles and laser beam incidence angles. The analysis primarily concentrates on conduction-mode laser melting, which forms the basis for a wide variety of welding and additive manufacturing techniques. Due to highly transient and complex molten metal flow behaviour, involving multiple physical effects such as solid-liquid phase transformation, laser-material interactions and free-surface oscillations, it is challenging to investigate the~evolution of melt-pool behaviour using merely trial-and-error experiments. In~the~present work, high-fidelity three-dimensional numerical simulations are performed to describe geometrical evolution and molten metal flow dynamics of the~melt pool. An~enhanced laser absorptivity model~\cite{Ebrahimi_2022} is utilised that takes into account the~effects of surface morphology, material composition and surface temperature on local energy absorption. The model has been rigorously validated with experimental data acquired using various laser beam shapes. Subsequently, the validated model is employed to predict the effects of different laser intensity profiles on the melt-pool behaviour and resulting solidification microstructure. The~present work offers a~simulation-based approach to predict the~effects of laser beam shaping in conduction-mode laser melting and to explore the~design space for process optimisation at a~reduced cost compared to experiments.

\FloatBarrier
\section{Methods} 
\label{sec:methods}

\begin{figure}[!htb] 
	\centering
	\includegraphics[width=1.0\linewidth]{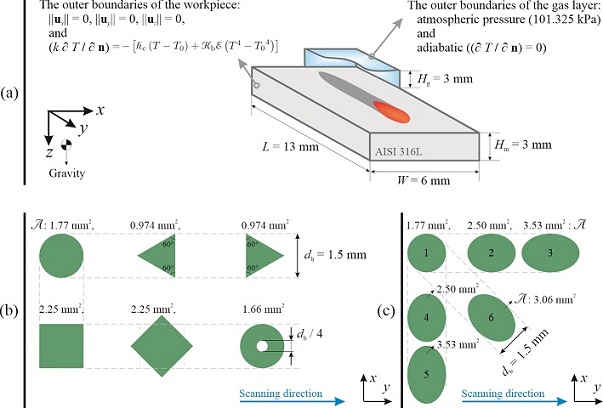}
	\caption{(a) Schematic of laser beam melting, plate dimensions and the~boundary conditions prescribed in numerical simulations. (b)~different beam shapes with uniform intensity profile (the~laser beam is perpendicular to the~$x$-$y$ plane), and (c)~different beam shapes obtained by inclining the~laser beam with (pseudo-)Gaussian intensity profile. Values shown above each laser beam shape indicate the~area of laser focal spot $\mathscr{A}$.}
	\label{fig:schematic}
\end{figure}

Three-dimensional physics-based numerical simulations were performed to study different beam shapes, intensity profiles and laser beam incidence angles in conduction-mode laser melting. The~geometrical configuration of the~problem, defined in a~Cartesian coordinate system, together with the~prescribed boundary conditions and the~beam shapes studied in the~present work are shown schematically in \cref{fig:schematic}. For beam shapes shown in \cref{fig:schematic}(b), a~uniform intensity profile was employed. Beam shapes shown in \cref{fig:schematic}(c) were produced by changing the~laser beam incidence angle about the~$x$ or $y$ axes, and the~beam had a~\mbox{(pseudo-)Gaussian} intensity profile. An~Yb:YAG laser with an~emission wavelength ($\lambda$) of $\SI{1.030e-6}{\meter}$, a~laser power ($\mathscr{P}$) of $\SI{700}{\watt}$ and a~focal spot diameter ($d_\mathrm{b}$) of $\SI{1.5e-3}{\meter}$ provides the~thermal energy flux required to locally melt the~plate that is made of stainless steel AISI~316L and is initially at room temperature ($T_\mathrm{i} = \SI{300}{\kelvin}$). The~laser beam scans the~surface at a~fixed travel speed ($\mathscr{V}$) of $\SI{1.5e-2}{\meter\per\second}$, unless stated otherwise. To~protect the~molten material from oxidation, an~argon shielding gas was employed. 

The~computational model was developed based on the~finite-volume approach using a~proprietary general-purpose solver, ANSYS~Fluent~\cite{Ansys_192}. The~volume-of-fluid~(VOF) method~\cite{Hirt_1981} was employed to model the~movement of metal-gas interface. Details of the~computational model are explained in our previous works~\cite{Ebrahimi_2022,Ebrahimi_2020,Ebrahimi_2019_conf,Ebrahimi_2022t}, and are not repeated here. In~the~present work, temperature-dependent material properties are employed for AISI~316L, and the~values are taken from~\cite{Ebrahimi_2022}. The~present model predicts multiple strongly-coupled physical processes that occur simultaneously during laser melting, such as variable laser absorptivity, molten metal flow due to Marangoni and thermal buoyancy forces, solid-liquid phase transformation and melt-pool surface deformation caused by the~violent internal flow and the~recoil pressure of metal vaporisation. Thermal and fluid flow fields and associated melt-pool surface movements were predicted by solving the~conservation equations of mass, momentum, energy and a~scalar field, called volume fraction~$\phi$ to capture the~position of the~melt-pool surface, assuming that both the~molten metal and argon are Newtonian fluids and that their densities are independent of any pressure variations that occur during processing.
These are defined as follows:

\begin{equation}
	\frac{D \rho}{D t} + \rho \left(\nabla \cdot \mathbf{u}\right) = 0,
	\label{eq:mass}
\end{equation}

\begin{equation}
	\rho \frac{D \mathbf{u}}{D t} = \mu \nabla^2 \mathbf{u} -\nabla p - C\ \frac{(1 - \psi)^2}{\psi^3 + \epsilon} \ \mathbf{u} + \mathbf{F}_\mathrm{s},
	\label{eq:momentum}
\end{equation}

\begin{equation}
	\rho \frac{D h}{D t} = \frac{k}{c_\mathrm{p}} \nabla^2 h - \rho \frac{D \left( \psi \mathcal{L}_\mathrm{f} \right)}{D t} + S_\mathrm{q} + S_\mathrm{l},
	\label{eq:energy}
\end{equation}

\begin{equation}
	\frac{D \phi}{D t} = 0.
	\label{eq:vof}
\end{equation}

\noindent
Here, $\rho$ is the~density, $\mathbf{u}$~the~fluid velocity vector, $t$~the~time, $\mu$~the~dynamic viscosity, $p$~the~pressure,  $C$~the~mushy-zone constant, equal to~$\SI[parse-numbers = false]{10^7}{\kilogram\per\square\meter\per\square\second}$~\cite{Ebrahimi_2019}, $\epsilon$~a~constant incorporated to avoid division by zero, equal to~$10^{-3}$, $h$~the~sensible heat, $k$~the~thermal conductivity, $c_\mathrm{p}$~the~specific heat capacity, $\mathcal{L}_\mathrm{f}$~the~latent heat of fusion, $\psi$~the~local liquid volume-fraction, and $\left( \psi \mathcal{L}_\mathrm{f} \right)$~the~latent~heat of the~material. The~value of $\phi$~in a~computational cell ranges between 0 and 1, representing the~volume-fraction of the~metal phase. Marangoni shear force, capillary force and recoil pressure are modelled by incorporating the~source term~$\mathbf{F}_\mathrm{s}$ into the~momentum equation (\cref{eq:momentum})~\cite{Ebrahimi_2022}. The~energy input from the~laser beam and heat losses from the~workpiece due to convection, radiation and vaporisation are modelled by adding the~source term $S_\mathrm{q}$ and the~sink term $S_\mathrm{l}$ to the~energy equation (\cref{eq:energy}) respectively~\cite{Ebrahimi_2022,Ebrahimi_2020}. The~boundary conditions prescribed at the~outer boundaries of the~computational domain are shown in~\cref{fig:schematic}. Spatial and temporal variations of the~laser absorptivity during the~process resulting from changes in surface temperature and laser beam incidence angle were modelled using an~enhanced absorption model that is described in our previous work~\cite{Ebrahimi_2022}. User-defined functions~(UDFs) were employed to implement the~source and sink terms in the~governing equations, the~laser absorptivity and surface tension models. The~accuracy of the~present computational model in predicting thermal profile and molten metal flow dynamics in laser melting is thoroughly investigated in our previous works~\cite{Ebrahimi_2019_conf,Ebrahimi_2020,Ebrahimi_2022,Ebrahimi_2022t} by comparing the~numerical results with experimental and theoretical data for different laser systems and various laser beam intensity profiles and shapes. The~validity of the~model is further investigated in the~present work, and the~results are presented in \cref{sec:validation}.

The~computational mesh consists of about~$\SI{2e6}{}$ non-uniform hexahedral cells with minimum cell spacing of~$\SI{4e-5}{\meter}$ close to the~gas-metal interface. The~computational cell spacing was chosen sufficiently fine to achieve at least $35$ cells along the~melt-pool width, which is sufficient to obtain grid-independent numerical predictions~\mbox{\cite{Ebrahimi_2020,Ebrahimi_2022,Ebrahimi_2022t}}. The~second-order central differencing scheme and the~first-order implicit formulation were employed for spatial and temporal discretisation respectively. An~explicit VOF method~\cite{Ubbink_1997} was employed to capture the~gas-metal interface. The~PISO~(pressure-implicit with splitting of operators) scheme~\cite{Issa_1986} was employed for resolving the~pressure-velocity coupling, and the~PRESTO~(pressure staggering option) scheme~\cite{Patankar_1980} was applied for the~pressure interpolation. To~keep the~value of the~Courant number \mbox{$(\mathrm{Co} = \lVert \mathbf{u} \rVert \Delta t / \Delta x)$} less than $0.25$, a~fixed time-step size of $\Delta t = \SI{1e-6}{\second}$ was employed. 

The~effects of laser beam shaping on microstructural grain structure are studied by computing the~values of solidification growth rate~($R$) and thermal gradient~($G$) at the~computational grid points located in the~solidifying region (\textit{i.e.} mushy region) as follows~\cite{Shi_2020}:

\begin{equation}
	G = \lVert \nabla T \rVert = \sqrt{\nabla T \cdot \nabla T},
	\label{eq:G}
\end{equation}

\begin{equation}
	R = \mathscr{V} \cdot \cos\left(\alpha\right).
	\label{eq:R}
\end{equation}

\noindent
Here, $\alpha$ is the~angle between the~maximum heat flow direction and the~scanning direction, and is determined as follows:

\begin{equation}
	\alpha = \arccos\left(\frac{\partial T / \partial y}{G}\right).
	\label{eq:solidification_angle}
\end{equation}

\noindent
The~values of solidification growth rate~($R$) and thermal gradient~($G$) obtained from the~simulations are plotted in the~so-called `solidification map'. The~solidification map is divided into three regions according to the~analytical model proposed by~\citet{Hunt_1984}, indicating the~nucleation propensity of fully~columnar (characterised by high $G/R$ values), fully~equiaxed (characterised by low $G/R$ values) and mixed columnar-equiaxed grains. According to the~Hunt's model, the~critical gradient conditions for fully columnar and fully equiaxed grain growth can be approximated respectively by

\begin{equation}
	G > \xi \, \Delta T_\mathrm{c} \, \sqrt[3]{100 \cdot N_0} \, \left(1 - \left(\frac{\Delta T_\mathrm{N}}{\Delta T_\mathrm{c}}\right)^3\right),
	\label{eq:hunt_columnar}
\end{equation}

\noindent
and

\begin{equation}
	G < \xi \, \Delta T_\mathrm{c} \, \sqrt[3]{N_0} \, \left(1 - \left(\frac{\Delta T_\mathrm{N}}{\Delta T_\mathrm{c}}\right)^3\right),
	\label{eq:hunt_equiaxed}
\end{equation}

\noindent
where, $\xi$ is a~constant equal to 0.617~\cite{Hunt_1984}, $\Delta T_\mathrm{c}$~the~undercooling for dendrite formation, $N_0$ the~nucleation density and $\Delta T_\mathrm{N}$~the~critical undercooling for nucleation. Assuming that the~tip velocity of the~dendrite is equal to the~solidification growth rate~$R$~\cite{Knapp_2019}, the~undercooling for dendrite formation is determined by an~empirical correlation as follows~\cite{Gaeumann_2001,Shi_2020}:

\begin{equation}
	\Delta T_\mathrm{c} = \left(\frac{R}{a}\right)^{1/b},
	\label{eq:DT_c}
\end{equation}

\noindent
where, $a$ and $b$ are constants whose values depend on the~material and are equal to $\SI{7.325e-6}{}$ and $\SI{3.12}{}$ respectively~\cite{Tan_2015} for the~alloy employed in the~present study. The~values of nucleation density $N_0$ and the~critical undercooling for nucleation~$\Delta T_\mathrm{N}$ are taken from the~literature~\cite{Tumkur_2021,Shi_2020,Roehling_2020} for stainless steel AISI~316L and are equal to $\SI[parse-numbers = false]{10^{15}}{\per\cubic\meter}$ and $\SI{10}{\kelvin}$ respectively.

\FloatBarrier
\section{Results and discussion}
\label{sec:results}

\subsection{Model validation}
\label{sec:validation} 
\begin{figure}[!htb] 
	\centering
	\includegraphics[width=1.0\linewidth]{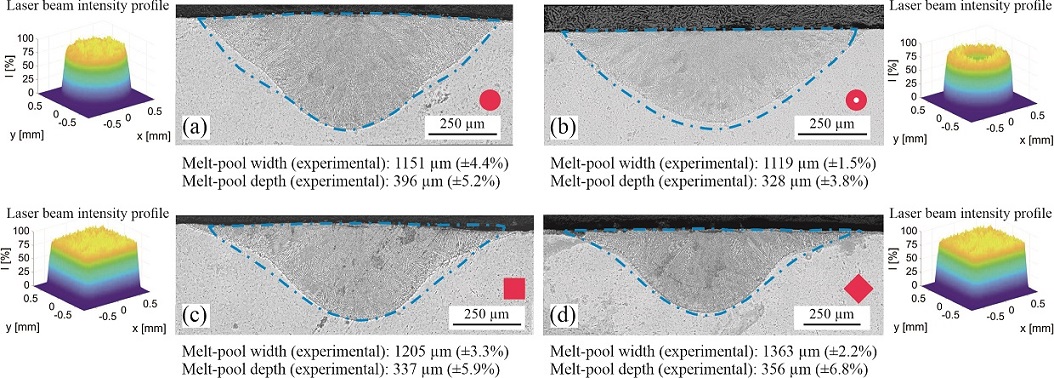}
	\caption{Comparison of the~numerically predicted melt-pool shapes using the~present computational model with the~experimental measurements for (a) circular, (b) annular, (c) square and (d) diamond laser beam shapes. Red symbols on each subfigure show the~laser beam shape. Blue curves show the~numerically predicted melt-pool shape.}
	\label{fig:validation}
\end{figure}

Four different laser beam shapes were studied to examine the~validity of the~present numerical simulations in predicting the~melt-pool shape and dimensions. An~Yb:YAG laser (Trumpf TruDisk 10001) with an~emission wavelength of $\lambda = \SI{1.030e-6}{\meter}$ in combination with three optical transport fibres (\textit{i.e.} a~$\SI{600}{\micro\meter}$ circular core fibre, a~$\SI{620}{\micro\meter}$ square core fibre and a~$400/100\,\SI{}{\micro\meter}$ 2-in-1 core fibre) was employed in the~experiments. The~focusing optics (BEO D70 Trumpf) consist of a~collimator with a~focal length of $\SI{2e-1}{\meter}$ and lenses with two different focal lengths of $\SI{4e-1}{\meter}$ and $\SI{6e-1}{\meter}$. Employing the~lens with a~focal length of $\SI{6e-1}{\meter}$ in combination with the~2-in-1 fibre, a~beam with a~focal spot size of $d_\mathrm{b} = \SI{1.2e-3}{\meter}$ and a~uniform intensity profile in the~focal plane, as shown in~\cref{fig:validation}, was produced. The~laser power ($\mathscr{P}$) was set to $\SI{500}{\watt}$ to locally melt a~stainless steel AISI~316L plate at a~fixed travel speed of $\mathscr{V} = \SI{2e-2}{\meter\per\second}$. The~plate employed in the~experiments had a~cuboid shape with dimensions ($x$, $y$, $z$) of $100\times250\times\SI{10}{\cubic\milli\meter}$. Each experimental trial was conducted with a minimum of three repetitions to establish the replicability of the outcomes. To capture the~experimental melt-pool shapes, macrographs were extracted \textit{ex-situ} from middle of the~melting track and photographed using a~digital microscope. Uncertainties in experimental measurements of the~melt-pool depth and width are about $4$--$7\%$ and $2$--$4\%$ respectively. Iso-surfaces of solidus temperature ($T_\mathrm{s} = \SI{1658}{\kelvin}$) obtained from the~numerical simulations were used to visualise the~melt-pool shape. A~comparison is made between the~numerically predicted and experimentally measured melt-pool shapes for different beams, and the~results are shown in~\cref{fig:validation}. The~maximum absolute error in the~prediction of melt-pool depth and width are less than $7\%$ and $5\%$ respectively, demonstrating a~reasonable agreement without exercising any parameter tuning. Such an~error is attributed to the~simplifying assumptions made to develop the~model, uncertainties associated with modelling material properties at elevated temperatures and experimental variations. 

\begin{figure}[!htb] 
	\centering
	\includegraphics[width=1.0\linewidth]{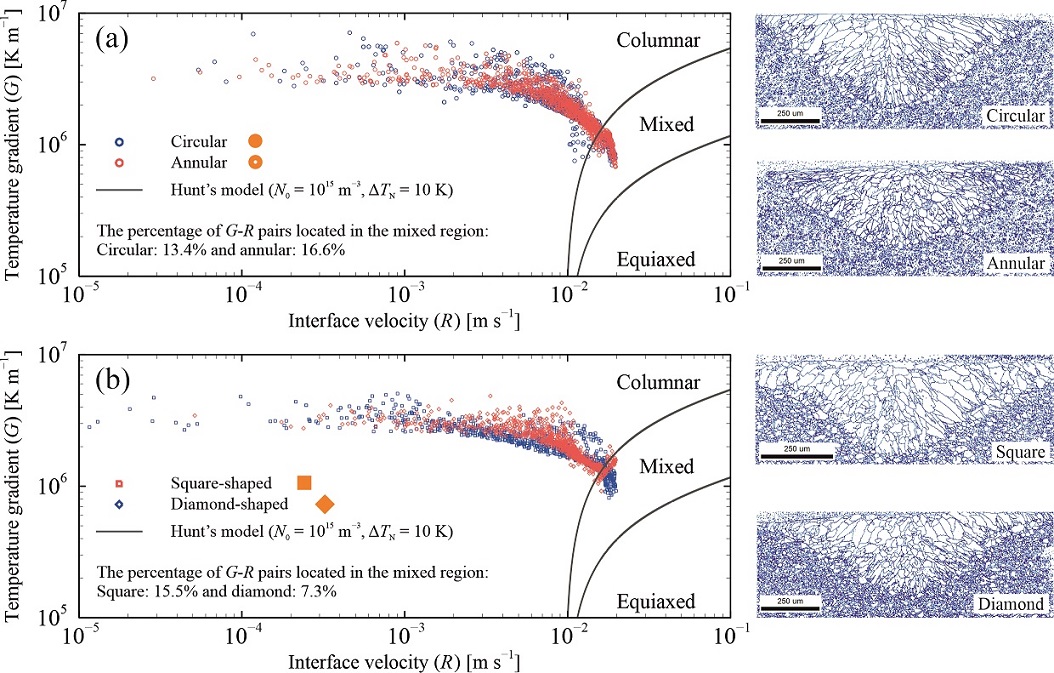}
	\caption{The distribution of numerically predicted grain morphologies on the solidification map based on Hunt's model (left column) and the EBSD measurements (right column) for different laser beam shapes. Orange symbols show the laser beam shape.}
	\label{fig:validation_ebsd}
\end{figure}

The~numerically predicted $G$-$R$ pairs for different laser beam shapes are plotted in the~solidification map, as shown in \cref{fig:validation_ebsd}. The~solidification map is divided into three regions according to the~analytical model proposed by~\mbox{\citet{Hunt_1984}}. EBSD (electron backscatter diffraction) technique was used to visualise microstructural grain structure on the~transverse cross-sections of the~melting track after the~experiments, and the~results are shown in~\cref{fig:validation_ebsd}. For EBSD measurements, the samples were further polished with colloidal silica for 50~minutes. A~Thermo Scientific Helios G4 PFIB UXe SEM equipped with an~EDAX detector was employed for EBSD analysis. A~specimen tilt angle of 60$^\circ$ along with an~accelerating voltage of $\SI{20}{\kilo\volt}$ and a~current of $\SI{3.2}{\nano\ampere}$ was used for data acquisition. A~step size of $\SI{0.8}{\micro\meter}$ was employed for the measurements. EBSD data processing was performed using EDAX OIM Analysis 8 software. The~examination of EBSD data shows that changes in beam shape result in variations in the~proportion of columnar grains. The~EBSD-derived percentages of columnar grains for circular, annular, square, and diamond-shaped beams are approximately 88\%, 85\%, 86\%, and 95\%, respectively. This observation demonstrates a~satisfactory agreement between the~numerically predicted grain morphologies following Hunt's model and the~EBSD-derived measurements.

\FloatBarrier
\subsection{Melt-pool behaviour}
\label{sec:shape}

\begin{figure}[htb] 
	\centering
	\includegraphics[width=1.0\linewidth]{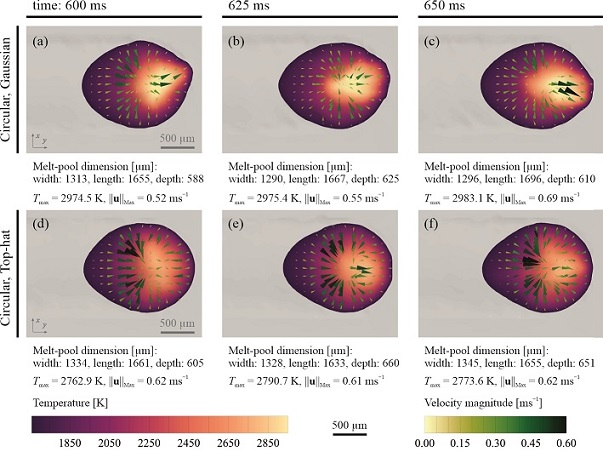}
	\caption{Snapshots of thermal and fluid flow fields over the~melt-pool surface at three different time instances and the~predicted melt-pool dimensions for circular laser beams with Gaussian (a-c) and top-hat (d-f) intensity profiles. The~laser power is $\SI{700}{\watt}$ and the~travel speed is $\SI{1.5e-2}{\meter\per\second}$. The~laser beam is perpendicular to the~plate surface.}
	\label{fig:tf_gth}
\end{figure}

\Cref{fig:tf_gth} shows three snapshots of thermal and fluid flow fields over the~melt-pool surface for circular beams with Gaussian and top-hat intensity profiles after reaching the~quasi-steady-state condition. The~results show that the~molten metal flows from the~melt-pool rim towards its central part because of the~Marangoni shear force, which agrees with experimental observations of \mbox{\citet{Mills_1998}}. The~molten metal temperature in the~central part of the~pool is above the~critical temperature for which the~value of the~temperature gradient of surface tension ($\mathrm{d}\gamma / \mathrm{d}T$) is positive, resulting in an~outward Marangoni flow from the~central part towards the~melt-pool rim. A~complex unsteady flow pattern forms in the~central part of the~pool where the~inward and the~outward streams interact. Although the~melt-pool resembles an~almost symmetric shape with respect to the~$y$-$z$ plane, the~molten metal flow and the~temperature distribution in the~pool are inherently unsteady and asymmetric due to hydrodynamic instabilities, requiring a~transient and fully three-dimensional model to simulate transport phenomena in the~pool properly. The~thermal energy absorbed by the~material diffuses through the~material by conduction and advects with fluid flow in the~pool. The~overall energy transfer due to heat conduction and advection in combination with heat loss from the~material determines the~melt-pool shape; however, heat loss from the~material due to radiation and convection is about $0.3\%$ of the~absorbed laser optical power, and thus its effect is negligible. Maximum fluid velocity in the~pool is in the~order of $\mathscr{O}(\SI{0.6}{\meter\per\second})$, resulting in a~P\'eclet number ($\mathrm{Pe} = \rho c_\mathrm{p} \mathscr{D} \lVert \mathbf{u} \rVert / k$) of about 100 that signifies the~dominant effect of advection in total energy transfer. 

For the~process parameters studied here, the~melt-pool dimensions obtained using circular beams with top-hat and Gaussian intensity profiles are comparable but the~peak temperature obtained using a~beam with a~Gaussian intensity profile is about $7\%$ higher than that using a~beam with a~top-hat intensity profile. This is because the~peak beam intensity of a~Gaussian profile is higher than that of a~top-hat profile for similar laser power and focal spot diameter. However, the~intensity of the~Gaussian beam rapidly reduces to values less than that of the~top-hat beam when moving away from the pool centre. The~greater peak temperature at the~pool centre obtained using the~Gaussian beam also results in higher absorptivity of laser energy that increases with temperature~\cite{Ebrahimi_2022}. Accordingly, the~magnitudes of temperature gradient and Marangoni shear force for the~case with a~Gaussian intensity profile are greater than those with a~top-hat intensity profile, resulting in more sensitivity to spatial disturbances when using a~beam with a~Gaussian intensity profile. The~maximum fluid velocity predicted for the~case with a~Gaussian intensity profile seems to be generally lower than that with a~top-hat intensity profile, which is attributed to the~greater outward Marangoni force in the~former case.

\begin{figure}[htb] 
	\centering
	\includegraphics[width=0.7\linewidth]{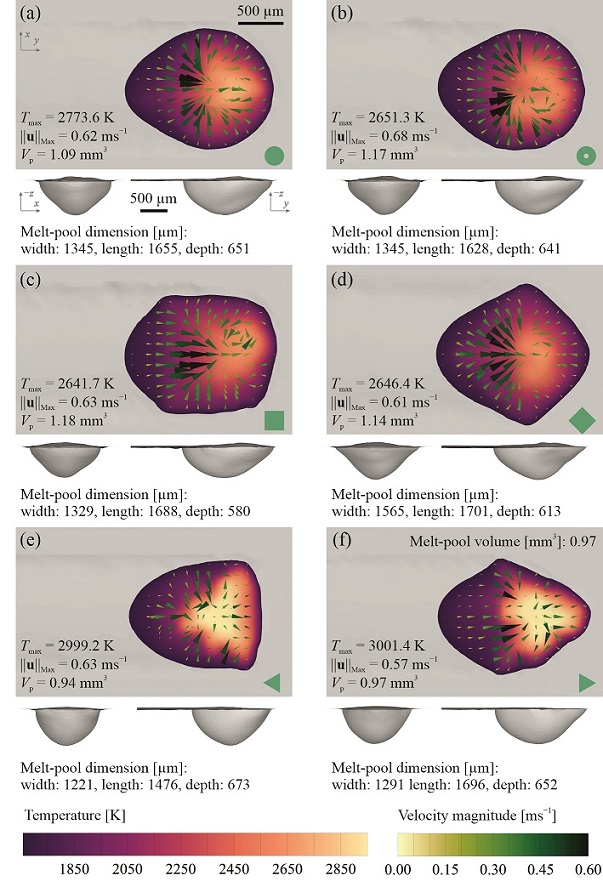}
	\caption{Thermal and fluid flow fields over the~melt-pool surface and the~predicted melt-pool shapes for different laser beam shapes after reaching the~quasi-steady-state condition ($t = \SI{0.65}{\second}$). The~laser power is $\SI{700}{\watt}$, the~beam has a~uniform intensity profile and the~travel speed is $\SI{1.5e-2}{\meter\per\second}$. Green symbols on each subfigure show the~laser beam shape.}
	\label{fig:tf_bsh}
\end{figure}

Melt-pool behaviour in laser melting cannot be characterised solely based on laser power, spot size, travel speed and material properties; and information about beam shape and laser intensity profile is also required. Reshaping the~laser beam can lead to changes in power density~$\mathscr{P}_\mathrm{d}$ and interaction time ($t_\mathrm{i}$), the~product of which results in energy density~$\mathscr{E}_\mathrm{d}$~\cite{Steen_2010} that critically affects the~melt-pool behaviour~\cite{Assuncao_2012}. \Cref{fig:tf_bsh} shows the~numerically predicted temperature distribution and velocity vectors over the~melt-pool surface and melt-pool shape for different laser beam shapes with uniform intensity profiles. For all the~beam shapes studied here, an~inward flow from the~melt-pool rim is observed over the~surface that collides with an~outward flow close to the~pool centre, the~interaction of which leads to the~formation of vortex flow over the~surface that enhances fluid mixing in the~pool. Although the~total heat input to the~material is almost identical for different beam shapes\footnote{The~value of laser absorptivity depends on surface temperature, material composition and laser beam incidence angle~\cite{Ebrahimi_2022t}. Such effects are accounted for in the~absorptivity model employed in the~present numerical simulations. Hence, the~total energy input to the~material is not necessarily the~same for different cases studied in the~present work.}, the~predicted melt-pool dimensions differ considerably. The~results show that there is not necessarily a~one-to-one relationship between energy density and melt-pool volume, hence employing the~concept of energy density is insufficient to explain changes in melt-pool dimensions obtained using different beam shapes. Assuming that the~average laser absorptivity is similar for different beam shapes, it can be argued, based on the~conservation of energy principle, that the~material's total enthalpy must be identical when employing different beam shapes because the~total heat input is the~same. Therefore, variations in the~thermal field should also be accounted for in addition to the energy density to explain the~difference between melt-pool shapes obtained using different beam shapes. Changes in the~pool shape also lead to changes in the~direction and magnitude of temperature gradients, affecting the~grain morphology of the~material after  solidification~\cite{Rappaz_1989}.

\begin{figure}[htb] 
	\centering
	\includegraphics[width=0.7\linewidth]{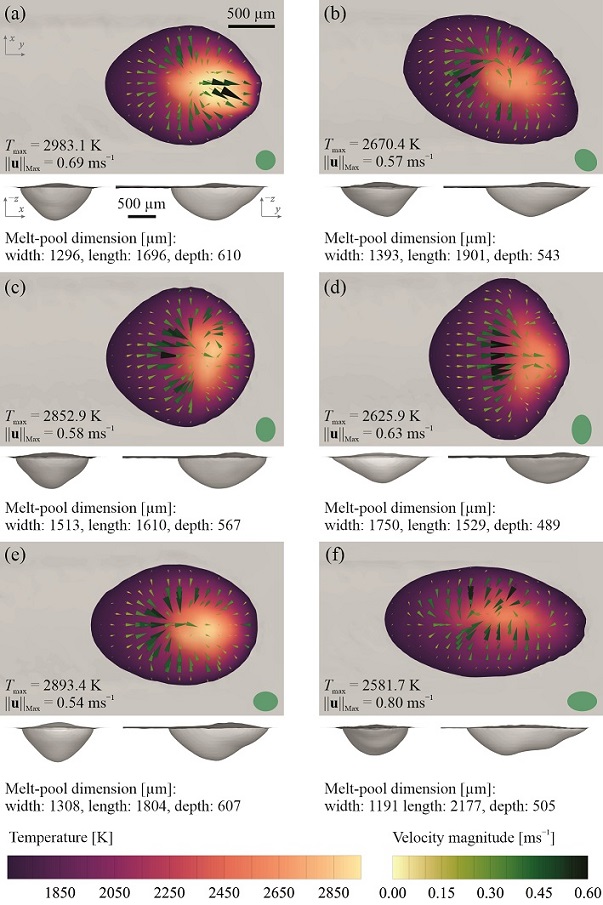}
	\caption{The~effects of laser beam incidence angle on thermal and fluid flow fields over the~melt-pool surface and the~pool shapes after reaching quasi-steady-state condition ($t = \SI{0.65}{\second}$). The~laser beam is perpendicular to the~$x$-$y$ plane in subfigure a. Beam shape in subfigure b is produced by inclining the~beam about both $x$ and $y$ axes by $45^\circ$. Beam shapes in subfigures c and d are produced by inclining the~beam about the~$y$-axis by $45^\circ$ and $60^\circ$ respectively. Similarly, beam shapes in subfigures e and f are produced by inclining the~beam about the~$x$-axis by $45^\circ$ and $60^\circ$ respectively. The~laser power is $\SI{700}{\watt}$, the~beam has a~(pseudo-)Gaussian intensity profile and the~travel speed is $\SI{1.5e-2}{\meter\per\second}$. Green symbols on each subfigure show the~laser beam shape.}
	\label{fig:tf_iang}
\end{figure}

Elliptical beam shapes with different energy densities can be produced by changing the~incidence angle of circular laser beams. A~change in the~laser beam incidence angle not only distorts the~beam shape but also affects the~laser absorptivity and hence the total energy input to the~material. \Cref{fig:tf_iang} shows the~numerically predicted temperature distribution and velocity vectors over the~melt-pool surface and the~pool shape for different laser beam incidence angles with (pseudo-)Gaussian intensity profiles. The~flow pattern obtained using inclined beams seems similar to that obtained using the~respective circular beam, and the~molten metal flows from the melt-pool rim toward its centre. The~melt-pool shape projected on the~$x$-$z$ plane resembles a~symmetric shape for all the~laser beam inclination angles except those where the~beam is inclined about both $x$ and $y$ axes (see \cref{fig:tf_iang}(b)). For the~cases studied in the~present work where an~Yb:YAG laser with an~emission wavelength of $\SI{1.030e-6}{\meter}$ is employed, the~average absorptivity of stainless steel AISI~316L increases by 2--3\% and 8--12\% by increasing the~inclination angle from $0^\circ$ to $45^\circ$ and $60^\circ$ respectively~\cite{Ebrahimi_2022}. Although total energy input to the~material increases by inclining the~laser beam due to enhanced laser absorptivity, power density decreases with increasing the~inclination angle because of the~considerable increase in the~spot area~$\mathscr{A}$ (41.4\% for an~inclination angle of $45^\circ$ and 99.4\% for an~inclination angle of $60^\circ$). When the~laser spot is elongated in the~transverse direction (\textit{i.e.} inclining the~beam about the~$y$-axis), the~interaction time remains the~same as that of the~circular beam; therefore, the~overall effect of inclining the~laser beam is a~decrease in the~energy density and thus the~melt-pool depth and peak temperature. Similarly, the~power density decreases when the~spot is elongated in the~scanning direction because of the~increased spot area; however, decrease in the~energy density is less than that of the~case with a~laser spot elongated in the~transverse direction due to the~increased interaction time. Hence, deeper melt-pools are observed when the~beam is elongated in the~scanning direction compared to those where the~spot is elongated in the~transverse direction. The~present numerical results agree with the~experimental observations reported in the~literature~\cite{Liao_2007,Ayoola_2019,Mi_2022,FathiHafshejani_2022} for conduction-mode laser melting. 

\FloatBarrier
\subsection{Bead profile}
\label{sec:bead_profile}

\begin{figure}[!htb] 
	\centering
	\includegraphics[width=0.6\linewidth]{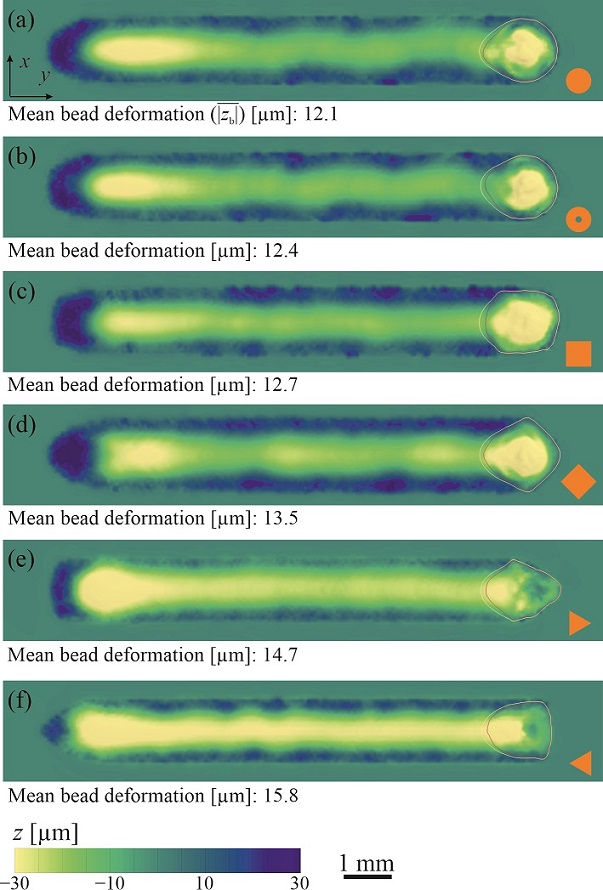}
	\caption{Surface profile of the~bead obtained from the~numerical simulations for different laser beam shapes ($t = \SI{0.65}{\second}$). The~laser power is $\SI{700}{\watt}$, the~beam has a~uniform intensity profile and the~travel speed is $\SI{1.5e-2}{\meter\per\second}$. Orange symbols on each subfigure show the~laser beam shape. Negative values of~$z$ indicate surface elevation. Bead deformation values over the melt track, excluding the molten region, were employed to compute mean bead deformations.}
	\label{fig:sp_bsh}
\end{figure}

Bead profile and surface quality after laser melting are influenced by the~thermal and fluid flow fields in the~melt-pool. \Cref{fig:sp_bsh}~shows the~surface profile of the~bead measured in the~$z$ direction with reference to the~un-melted surface and the~arithmetic mean of the~absolute bead deformation ($\overline{\left|z_\mathrm{b}\right|}$) for different laser beam shapes with uniform intensity profiles. The~bead surface is elevated in the~central region of the~melt track due to inward molten metal flow during laser melting for all the beam shapes shown in~\cref{fig:sp_bsh}. Since the~mass of the~workpiece is almost unchanged during the~process (\textit{i.e.}~vaporisation of the~material and changes in the~material density are negligible in conduction-mode laser melting), the~bead surface is depressed in regions close to the~outer boundaries of the~track. The~mean bead deformations obtained using circular, annular and square beam shapes appear to be similar and in the~order of $\SI{12}{\micro\meter}$ that is lower than those obtained using diamond and triangular beam shapes, which can be attributed to the~lower energy densities of the~former cases. For some cases, the~crest trail of the~bead shows a~wavy pattern that is due to molten metal flow instabilities during laser melting. Although no humping defect is observed for the~cases studied in the~present work, the~surface profile obtained using the~diamond beam shape (\cref{fig:sp_bsh}(d)) suggests that the~bead could be prone to a~humping defect.

\begin{figure}[!htb] 
	\centering
	\includegraphics[width=0.6\linewidth]{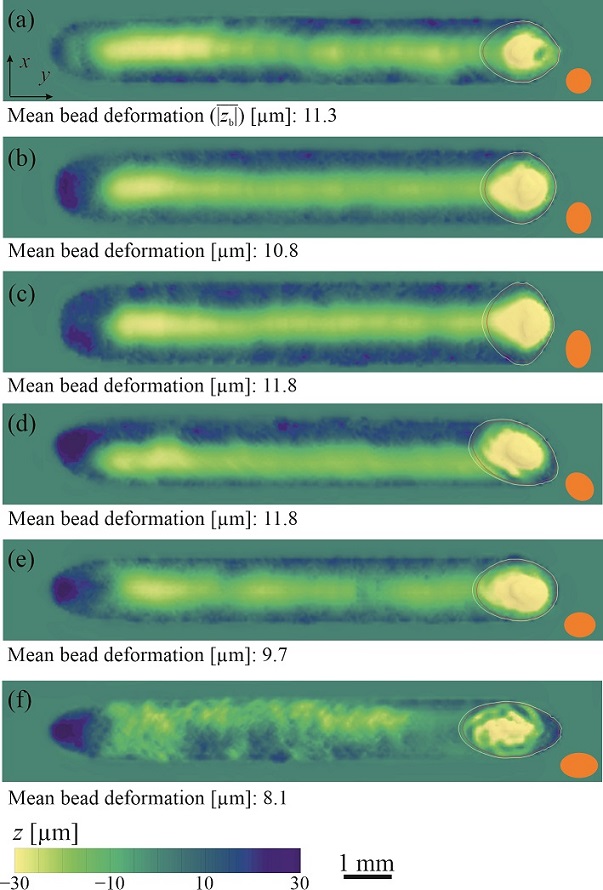}
	\caption{Surface profile of the~bead obtained from the~numerical simulations for different laser beam incidence angles ($t = \SI{0.65}{\second}$). The~laser power is $\SI{700}{\watt}$, the~beam has a~(pseudo-)Gaussian intensity profile and the~travel speed is $\SI{1.5e-2}{\meter\per\second}$. Orange symbols on each subfigure show the~laser beam shape. Negative values of~$z$ indicate surface elevation. Bead deformation values over the melt track, excluding the molten region, were employed to compute mean bead deformations.}
	\label{fig:sp_iang}
\end{figure}

\Cref{fig:sp_iang} shows the~surface profile of the~bead for different laser beam shapes with \mbox{(pseudo-)Gaussian} intensity profiles. The~bead surface produced using the~circular beam with a~Gaussian intensity profile shows solidified waves that result from pulsating molten metal flow instabilities~\cite{Ebrahimi_2022t,Ebrahimi_2020,Kidess_2016_PhysFlu}. Elongation of the~laser spot in the~transverse direction increases the~bead width and seems to have an~insignificant influence on the~mean bead deformation. In contrast, elongation of the~laser spot in the~scanning direction does not change the~bead width and reduces the~mean bead deformation. The results indicate that inclining the~beam about the~$y$-axis by $60^\circ$ (\cref{fig:sp_iang}(f)) results in irregular surface deformations that can be ascribed to the flow instabilities in the melt pool. The~bead profile obtained using a~beam inclined about both $x$ and $y$ axes (\cref{fig:sp_iang}(d)) differs from the~other cases and is elevated in regions close to one edge of the~melt track and is depressed in regions close to the~other edge. The~results show that the~laser beam inclination angle can be utilised as a~potential means to control bead profile in laser melting processes.

\FloatBarrier
\subsection{Microstructural grain morphology}
\label{sec:texture}

\begin{figure}[!htb] 
	\centering
	\includegraphics[width=0.7\linewidth]{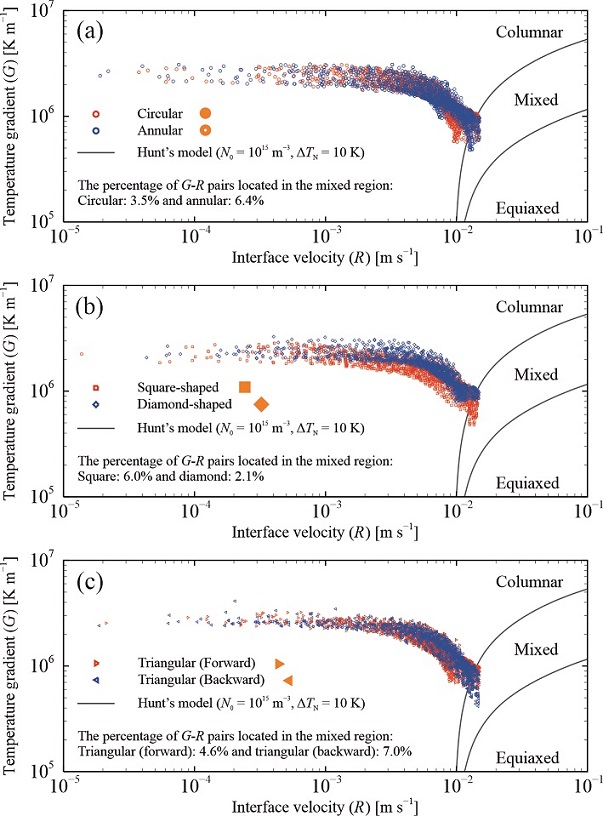}
	\caption{Scatter plot of $G$-$R$ data on the~solidification map for different laser beam shapes with uniform intensity profiles after reaching quasi-steady-state condition ($t = \SI{0.65}{\second}$). The~laser power is $\SI{700}{\watt}$ and the~travel speed is $\SI{1.5e-2}{\meter\per\second}$. Orange symbols show the~laser beam shape.}
	\label{fig:hm_bsh}
\end{figure}

\begin{figure}[!htb] 
	\centering
	\includegraphics[width=0.7\linewidth]{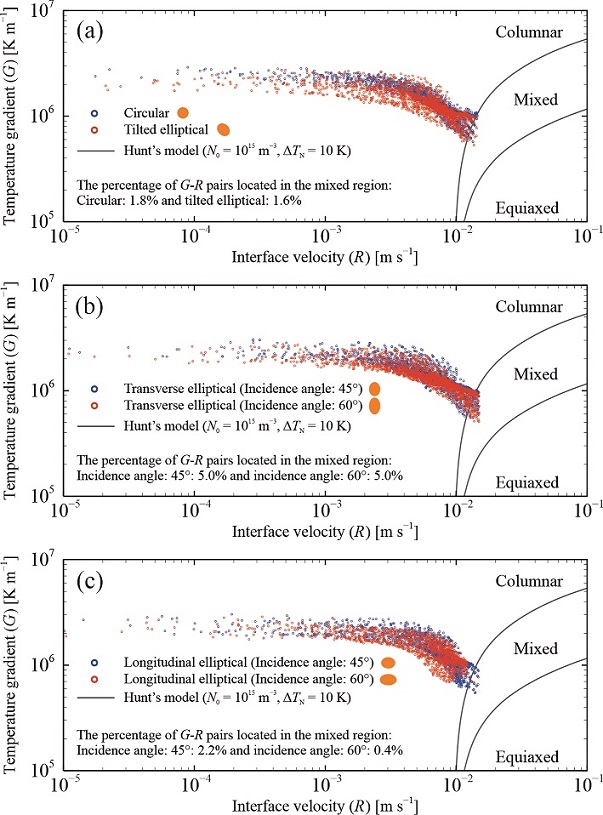}
	\caption{Scatter plot of $G$-$R$ data on the~solidification map for different laser beam inclination angles with Gaussian intensity profiles after reaching quasi-steady-state condition ($t = \SI{0.65}{\second}$). The~laser power is $\SI{700}{\watt}$ and the~travel speed is $\SI{1.5e-2}{\meter\per\second}$. Orange symbols show the~laser beam shape.}
	\label{fig:hm_iang}
\end{figure}

The~present high-fidelity numerical predictions revealed that laser beam shaping significantly affects the~melt-pool shape and thermal and fluid flow fields in conduction-mode laser melting. This can lead to changes in the~material's microstructural grain structure and macroscopic properties after solidification. The~distribution of $G$-$R$ pairs on the~solidification map are shown in~\cref{fig:hm_bsh,fig:hm_iang} for different laser beam shapes and inclination angles respectively. The results suggest that changing the~laser beam shape leads to the~variation of the~ratio of columnar to equiaxed grains. Since the~travel speed was the~same in all the~simulations, the~value of solidification growth rate $R$, which scales with the~travel speed (see~\cref{eq:R}), is similar for all the~cases. However, the~value of temperature gradient $G$ ranges between $\SI{4e5}{\kelvin\per\meter}$ and $\SI{7e6}{\kelvin\per\meter}$ for different laser beam shapes and inclination angles studied in the~present work. Due to the~large magnitude of temperature gradients in conduction-mode laser melting ($\mathscr{O}(10^6)$~\si{\kelvin}), it is unlikely that equiaxed grains nucleate homogeneously during solidification~\cite{Kurz_2001,He_2021}. Grid points located at the~melt-pool tail on the~track centreline are the~farthest points from the~centroid of the~laser spot and have the~lowest temperature gradient~$G$ and the~highest solidification growth rate~$R$ (\textit{i.e.}~the~lowest~$G/R$); hence, equiaxed and mixed columnar-equiaxed grains are more likely to form in that region. For partially penetrated melt-pools, the~highest~$G/R$ values are found at the~fusion boundary and close to the~melt-pool bottom and sides since the~solidification growth rate~$R$ is minimum at those locations; thus, non-equiaxed (\textit{i.e.}~planar, cellular and columnar) grains are expected to form in those regions~\cite{Roehling_2017}. 

The~results presented in~\cref{fig:hm_bsh} indicate that changing the~beam shape from circular shape leads to a~decrease in the~nucleation propensity of fully-columnar grains except for the diamond beam shape. Calculating the~percentage of $G$-$R$ pairs located in the~mixed columnar-equiaxed region for different laser beam shapes, it is found that employing backward triangular~(7.0\%), annular~(6.4\%) and square~(6.0\%) beam shapes results in the~highest nucleation propensity of equiaxed grains. The~distribution of $G$-$R$ pairs in~\cref{fig:hm_iang} indicate that elongation of the~laser spot in the~transverse direction decreases the~nucleation propensity of fully-columnar grains that can be ascribed to the~decrease in the~magnitude of temperature gradient $G$ at the~melt-pool tail. Elongation of laser spot in the~scanning direction by inclining the~beam by $45^\circ$ seems to have insignificant effect on the~nucleation propensity of fully-columnar grains; however further increase in the~inclination angle to $60^\circ$ leads to an~increase in the~nucleation propensity of fully-columnar grains. This is because melt-pool elongation in the~scanning direction results in an~increase in the~average angle between the~maximum heat flow direction and the~scanning direction ($\alpha$), leading to a~decrease in the~solidification growth rate ($R$) and an~increase in the~value of $G/R$.

\FloatBarrier
\section{Conclusions}
\label{sec:conclusions}
The~effects of laser beam shaping on melt-pool behaviour, solidified bead profile and microstructural gain morphology were studied comprehensively through high-fidelity numerical experiments for conduction-mode laser melting of stainless steel AISI~316L. Three-dimensional unsteady numerical simulations were performed to examine the~effects of laser beam intensity profile, shape and inclination angle on the~melt-pool behaviour. Critical physical phenomena in laser melting such as temporal and spatial variations of absorptivity, heat and fluid flow dynamics, solidification and melting, and free-surface oscillations were accounted for in the~present computational model. Moreover, experiments were performed using different laser beam shapes and the~validity of the~present numerical predictions was demonstrated. Based on the~results of the~present work, the~following conclusions are drawn.

\begin{itemize}
	\item Altering the~laser intensity profile leads to changes in the~thermal and fluid flow fields in the~melt pool, which in turn affect the~melt-pool dimensions, melt-track bead profile and grain morphology. Moreover, the~sensitivity of melt-pool responses to spatial and temporal disturbances, and hence process stability, depend on the~laser-intensity profile.
	
	\item For identical processing parameters, reshaping the~laser beam can change the~energy density that affects the~melt-pool behaviour. However, variations in the~melt-pool shape and volume caused by laser beam shaping cannot be described solely based on the~concept of energy density, and the~material's total enthalpy should be taken into consideration.
	
	\item Inclining the~laser beam not only distorts the~beam shape but also affects the~laser absorptivity and hence total energy input to the~material. However, the~increase in the~absorptivity due to laser beam inclination is considerably less than the~increase in the~spot area; thus, the~overall effect of laser beam inclination is a~reduction in the~power density. Employing identical processing parameters and inclination angles, a~deeper melt-pool was observed when the~laser spot was elongated in the~scanning direction compared to the~case in which the~spot was elongated in the~transverse direction, which can be ascribed to the~increased interaction time in the~former case.
	
	\item Laser beam shaping has a~notable effect on melt-track bead profile and can be utilised as a~potential means to control bead profile in laser welding and additive manufacturing.
	
	\item Laser beam shaping significantly affects microstructural grain morphology by influencing the~thermal gradient and solidification growth rate. The~present findings demonstrate that the~material's microstructure, and thus its macroscopic properties, can be adjusted locally through laser beam shaping.
\end{itemize}

The~findings of the~present work provide an~improved insight into the~effects of laser beam shaping in laser welding and additive manufacturing. The~simulation-based approach developed in the~present work offers new routes for design space exploration and process optimisation.

Future research in this area could focus on advancing the optimisation of laser beam shaping for laser-based additive manufacturing. One promising avenue is to explore multi-objective optimisation techniques that consider not only the~melt-pool behaviour and bead profile but also the microstructural grain morphology. This would enable the~development of tailor-made microstructures with desired properties for specific applications. Data-driven approaches, combining experimental data with numerical simulations, could be employed to create predictive models that guide the selection of optimal laser beam intensity profiles for achieving precise control over microstructural characteristics. Additionally, investigating the influence of various process parameters and materials on the interaction between laser beam shaping and microstructure could open new avenues for tailoring material properties in additive manufacturing, enhancing the design and performance of additively manufactured components.

\section*{Acknowledgement}
\label{sec:acknowledgement}

This research was carried out under project number S17024 in the~framework of the~Partnership Program of the~Materials innovation institute M2i (www.m2i.nl) and the~Netherlands Organisation for Scientific Research (www.nwo.nl). This research project is part of the~AiM2XL program (www.m2i.nl/aim2xl).

\section*{Author Contributions}
\label{sec:author_contributions}

Conceptualisation, A.E.; methodology, A.E.; software, A.E.; validation, A.E., A.B., A.S. and M.S.; formal analysis, A.E.; investigation, A.E.; resources, A.E., G.R.B.E.R. and M.J.M.H.; data curation,~A.E.; writing---original draft preparation, A.E.; writing---review and editing, A.E., A.B., M.S., G.R.B.E.R. and M.J.M.H.; visualisation, A.E.; project administration, A.E. and M.J.M.H.; and funding acquisition, G.R.B.E.R. and M.J.M.H.

\section*{Conflict of interest}
\label{sec:conflict_of_interest}

The~authors declare no conflict of interest.

\section*{Data availability}
\label{sec:data_availability}

The~raw/processed data required to reproduce these findings cannot be shared at this time due to their large size, but representative samples of the~research data are presented in the~paper. Other datasets generated during this study are available from the~corresponding author on reasonable request.

{
	\small
	\onehalfspacing

}
\bibliographystyle{article-bibstyle} 

\begin{thebibliography}{53}
		\providecommand{\natexlab}[1]{#1}
		\providecommand{\url}[1]{\texttt{#1}}
		\expandafter\ifx\csname urlstyle\endcsname\relax
		\providecommand{\doi}[1]{doi: #1}\else
		\providecommand{\doi}{doi: \begingroup \urlstyle{rm}\Url}\fi
		
		\bibitem[Herzog \textit{et~al.}(2016)Herzog, Seyda, Wycisk, and
		Emmelmann]{Herzog_2016}
		Herzog, D., Seyda, V., Wycisk, E., and Emmelmann, C.
		\newblock Additive manufacturing of metals.
		\newblock \emph{Acta Materialia}, 117:\penalty0 371--392, 2016.
		\newblock \doi{10.1016/j.actamat.2016.07.019}.
		
		\bibitem[Khare \textit{et~al.}(2007)Khare, Kaul, Ganesh, Kumar, Jagdheesh, and
		Nath]{Khare_2007}
		Khare, J., Kaul, R., Ganesh, P., Kumar, H., Jagdheesh, R., and Nath, A.~K.
		\newblock Laser beam shaping for microstructural control during laser surface
		melting.
		\newblock \emph{Journal of Laser Applications}, 19\penalty0 (1):\penalty0 1--7,
		2007.
		\newblock \doi{10.2351/1.2402522}.
		
		\bibitem[Roehling \textit{et~al.}(2017)Roehling, Wu, Khairallah, Roehling,
		Soezeri, Crumb, and Matthews]{Roehling_2017}
		Roehling, T.~T., Wu, S.~S., Khairallah, S.~A., Roehling, J.~D., Soezeri, S.~S.,
		Crumb, M.~F., and Matthews, M.~J.
		\newblock Modulating laser intensity profile ellipticity for microstructural
		control during metal additive manufacturing.
		\newblock \emph{Acta Materialia}, 128:\penalty0 197--206, 2017.
		\newblock \doi{10.1016/j.actamat.2017.02.025}.
		
		\bibitem[Roehling \textit{et~al.}(2020)Roehling, Shi, Khairallah, Roehling,
		Guss, McKeown, and Matthews]{Roehling_2020}
		Roehling, T.~T., Shi, R., Khairallah, S.~A., Roehling, J.~D., Guss, G.~M.,
		McKeown, J.~T., and Matthews, M.~J.
		\newblock Controlling grain nucleation and morphology by laser beam shaping in
		metal additive manufacturing.
		\newblock \emph{Materials {\&} Design}, 195:\penalty0 109071, 2020.
		\newblock \doi{10.1016/j.matdes.2020.109071}.
		
		\bibitem[Shi \textit{et~al.}(2020)Shi, Khairallah, Roehling, Heo, McKeown, and
		Matthews]{Shi_2020}
		Shi, R., Khairallah, S.~A., Roehling, T.~T., Heo, T.~W., McKeown, J.~T., and
		Matthews, M.~J.
		\newblock Microstructural control in metal laser powder bed fusion additive
		manufacturing using laser beam shaping strategy.
		\newblock \emph{Acta Materialia}, 184:\penalty0 284--305, 2020.
		\newblock \doi{10.1016/j.actamat.2019.11.053}.
		
		\bibitem[Mei \textit{et~al.}(2017)Mei, Yan, Chen, Wang, and Chen]{Mei_2017}
		Mei, L., Yan, D., Chen, G., Wang, Z., and Chen, S.
		\newblock Influence of laser beam incidence angle on laser lap welding quality
		of galvanized steels.
		\newblock \emph{Optics Communications}, 402:\penalty0 147--158, 2017.
		\newblock \doi{10.1016/j.optcom.2017.05.032}.
		
		\bibitem[Grünewald \textit{et~al.}(2021)Grünewald, Gehringer, Schmöller, and
		Wudy]{Gruenewald_2021}
		Grünewald, J., Gehringer, F., Schmöller, M., and Wudy, K.
		\newblock Influence of ring-shaped beam profiles on process stability and
		productivity in laser-based powder bed fusion of {AISI} {316L}.
		\newblock \emph{Metals}, 11\penalty0 (12):\penalty0 1989, 2021.
		\newblock \doi{10.3390/met11121989}.
		
		\bibitem[Fathi-Hafshejani \textit{et~al.}(2022)Fathi-Hafshejani,
		Soltani-Tehrani, Shamsaei, and Mahjouri-Samani]{FathiHafshejani_2022}
		Fathi-Hafshejani, P., Soltani-Tehrani, A., Shamsaei, N., and Mahjouri-Samani,
		M.
		\newblock Laser incidence angle influence on energy density variations, surface
		roughness, and porosity of additively manufactured parts.
		\newblock \emph{Additive Manufacturing}, 50:\penalty0 102572, 2022.
		\newblock \doi{10.1016/j.addma.2021.102572}.
		
		\bibitem[Ebrahimi \textit{et~al.}(2021{\natexlab{a}})Ebrahimi, Kleijn, and
		Richardson]{Ebrahimi_2021}
		Ebrahimi, A., Kleijn, C.~R., and Richardson, I.~M.
		\newblock A simulation-based approach to characterise melt-pool oscillations
		during gas tungsten arc welding.
		\newblock \emph{International Journal of Heat and Mass Transfer}, 164:\penalty0
		120535, 2021{\natexlab{a}}.
		\newblock \doi{10.1016/j.ijheatmasstransfer.2020.120535}.
		
		\bibitem[Ebrahimi \textit{et~al.}(2021{\natexlab{b}})Ebrahimi, Kleijn, Hermans,
		and Richardson]{Ebrahimi_2021_b}
		Ebrahimi, A., Kleijn, C.~R., Hermans, M. J.~M., and Richardson, I.~M.
		\newblock The effects of process parameters on melt-pool oscillatory behaviour
		in gas tungsten arc welding.
		\newblock \emph{Journal of Physics D: Applied Physics}, 54\penalty0
		(27):\penalty0 275303, 2021{\natexlab{b}}.
		\newblock \doi{10.1088/1361-6463/abf808}.
		
		\bibitem[Cooke \textit{et~al.}(2020)Cooke, Ahmadi, Willerth, and
		Herring]{Cooke_2020}
		Cooke, S., Ahmadi, K., Willerth, S., and Herring, R.
		\newblock Metal additive manufacturing: {Technology}, metallurgy and modelling.
		\newblock \emph{Journal of Manufacturing Processes}, 57:\penalty0 978--1003,
		2020.
		\newblock \doi{10.1016/j.jmapro.2020.07.025}.
		
		\bibitem[Ebrahimi \textit{et~al.}(2021{\natexlab{c}})Ebrahimi, Kleijn, and
		Richardson]{Ebrahimi_2020}
		Ebrahimi, A., Kleijn, C.~R., and Richardson, I.~M.
		\newblock Numerical study of molten metal melt pool behaviour during
		conduction-mode laser spot melting.
		\newblock \emph{Journal of Physics D: Applied Physics}, 54:\penalty0 105304,
		2021{\natexlab{c}}.
		\newblock \doi{10.1088/1361-6463/abca62}.
		
		\bibitem[Ebrahimi \textit{et~al.}(2022)Ebrahimi, Sattari, Bremer, Luckabauer,
		willem R.~B. E.~Römer, Richardson, Kleijn, and Hermans]{Ebrahimi_2022}
		Ebrahimi, A., Sattari, M., Bremer, S. J.~L., Luckabauer, M., willem R.~B.
		E.~Römer, G., Richardson, I.~M., Kleijn, C.~R., and Hermans, M. J.~M.
		\newblock The influence of laser characteristics on internal flow behaviour in
		laser melting of metallic substrates.
		\newblock \emph{Materials {\&} Design}, 214:\penalty0 110385, 2022.
		\newblock \doi{10.1016/j.matdes.2022.110385}.
		
		\bibitem[Ebrahimi and Hermans(2023)]{Ebrahimi_2023}
		Ebrahimi, A. and Hermans, M.~J.
		\newblock Laser butt welding of thin stainless steel {316L} sheets in
		asymmetric configurations: {A} numerical study.
		\newblock \emph{Journal of Advanced Joining Processes}, 8:\penalty0 100154,
		2023.
		\newblock \doi{10.1016/j.jajp.2023.100154}.
		
		\bibitem[Ayoola \textit{et~al.}(2019)Ayoola, Suder, and Williams]{Ayoola_2019}
		Ayoola, W.~A., Suder, W.~J., and Williams, S.~W.
		\newblock Effect of beam shape and spatial energy distribution on weld bead
		geometry in conduction welding.
		\newblock \emph{Optics {\&} Laser Technology}, 117:\penalty0 280--287, 2019.
		\newblock \doi{10.1016/j.optlastec.2019.04.025}.
		
		\bibitem[Tenbrock \textit{et~al.}(2020)Tenbrock, Fischer, Wissenbach,
		Schleifenbaum, Wagenblast, Meiners, and Wagner]{Tenbrock_2020}
		Tenbrock, C., Fischer, F.~G., Wissenbach, K., Schleifenbaum, J.~H., Wagenblast,
		P., Meiners, W., and Wagner, J.
		\newblock Influence of keyhole and conduction mode melting for top-hat shaped
		beam profiles in laser powder bed fusion.
		\newblock \emph{Journal of Materials Processing Technology}, 278:\penalty0
		116514, 2020.
		\newblock \doi{10.1016/j.jmatprotec.2019.116514}.
		
		\bibitem[Tumkur \textit{et~al.}(2021)Tumkur, Voisin, Shi, Depond, Roehling, Wu,
		Crumb, Roehling, Guss, Khairallah, and Matthews]{Tumkur_2021}
		Tumkur, T.~U., Voisin, T., Shi, R., Depond, P.~J., Roehling, T.~T., Wu, S.,
		Crumb, M.~F., Roehling, J.~D., Guss, G., Khairallah, S.~A., and Matthews,
		M.~J.
		\newblock Nondiffractive beam shaping for enhanced optothermal control in metal
		additive manufacturing.
		\newblock \emph{Science Advances}, 7\penalty0 (38), 2021.
		\newblock \doi{10.1126/sciadv.abg9358}.
		
		\bibitem[Mi \textit{et~al.}(2022)Mi, Mahade, Sikström, Choquet, Joshi, and
		Ancona]{Mi_2022}
		Mi, Y., Mahade, S., Sikström, F., Choquet, I., Joshi, S., and Ancona, A.
		\newblock Conduction mode laser welding with beam shaping using a deformable
		mirror.
		\newblock \emph{Optics {\&} Laser Technology}, 148:\penalty0 107718, 2022.
		\newblock \doi{10.1016/j.optlastec.2021.107718}.
		
		\bibitem[Pamarthi \textit{et~al.}(2023)Pamarthi, Sun, Das, and
		Franciosa]{Pamarthi_2023}
		Pamarthi, V.~V., Sun, T., Das, A., and Franciosa, P.
		\newblock Tailoring the weld microstructure to prevent solidification cracking
		in remote laser welding of {AA}6005 aluminium alloys using adjustable
		ringmode beam.
		\newblock \emph{Journal of Materials Research and Technology}, 25:\penalty0
		7154--7168, 2023.
		\newblock \doi{10.1016/j.jmrt.2023.07.154}.
		
		\bibitem[Cloots \textit{et~al.}(2016)Cloots, Uggowitzer, and
		Wegener]{Cloots_2016}
		Cloots, M., Uggowitzer, P.~J., and Wegener, K.
		\newblock Investigations on the microstructure and crack formation of {IN738LC}
		samples processed by selective laser melting using {Gaussian} and doughnut
		profiles.
		\newblock \emph{Materials {\&} Design}, 89:\penalty0 770--784, 2016.
		\newblock \doi{10.1016/j.matdes.2015.10.027}.
		
		\bibitem[Kubiak \textit{et~al.}(2015)Kubiak, Piekarska, and Stano]{Kubiak_2015}
		Kubiak, M., Piekarska, W., and Stano, S.
		\newblock Modelling of laser beam heat source based on experimental research of
		{Yb}:{YAG} laser power distribution.
		\newblock \emph{International Journal of Heat and Mass Transfer}, 83:\penalty0
		679--689, 2015.
		\newblock \doi{10.1016/j.ijheatmasstransfer.2014.12.052}.
		
		\bibitem[Collins \textit{et~al.}(2016)Collins, Brice, Samimi, Ghamarian, and
		Fraser]{Collins_2016}
		Collins, P.~C., Brice, D., Samimi, P., Ghamarian, I., and Fraser, H.
		\newblock Microstructural control of additively manufactured metallic
		materials.
		\newblock \emph{Annual Review of Materials Research}, 46\penalty0 (1):\penalty0
		63--91, 2016.
		\newblock \doi{10.1146/annurev-matsci-070115-031816}.
		
		\bibitem[Bremer \textit{et~al.}(2023)Bremer, Luckabauer, and willem
		R.B.E.~Römer]{Bremer_2023}
		Bremer, S.~J., Luckabauer, M., and willem R.B.E.~Römer, G.
		\newblock Laser intensity profile as a means to steer microstructure of
		deposited tracks in {Directed Energy Deposition}.
		\newblock \emph{Materials {\&} Design}, 227:\penalty0 111725, 2023.
		\newblock \doi{10.1016/j.matdes.2023.111725}.
		
		\bibitem[Dickey \textit{et~al.}(2018)Dickey, Holswade, and Shealy]{Dickey_2018}
		Dickey, F.~M., Holswade, S.~C., and Shealy, D.~L., editors.
		\newblock \emph{Laser Beam Shaping Applications}.
		\newblock {CRC} Press, 2018.
		\newblock \doi{10.1201/9781420028065}.
		
		\bibitem[Salter and Booth(2019)]{Salter_2019}
		Salter, P.~S. and Booth, M.~J.
		\newblock Adaptive optics in laser processing.
		\newblock \emph{Light: Science {\&} Applications}, 8\penalty0 (1), 2019.
		\newblock \doi{10.1038/s41377-019-0215-1}.
		
		\bibitem[Bi \textit{et~al.}(2023)Bi, Wu, Li, Yang, Jia, Starostenkov, and
		Dong]{Bi_2023}
		Bi, J., Wu, L., Li, S., Yang, Z., Jia, X., Starostenkov, M.~D., and Dong, G.
		\newblock Beam shaping technology and its application in metal laser additive
		manufacturing: A review.
		\newblock \emph{Journal of Materials Research and Technology}, 26:\penalty0
		4606--4628, 2023.
		\newblock \doi{10.1016/j.jmrt.2023.08.037}.
		
		\bibitem[Sundqvist \textit{et~al.}(2016)Sundqvist, Kaplan, Shachaf, Brodsky,
		Kong, Blackburn, Assuncao, and Quintino]{Sundqvist_2016}
		Sundqvist, J., Kaplan, A. F.~H., Shachaf, L., Brodsky, A., Kong, C., Blackburn,
		J., Assuncao, E., and Quintino, L.
		\newblock Numerical optimization approaches of single-pulse conduction laser
		welding by beam shape tailoring.
		\newblock \emph{Optics and Lasers in Engineering}, 79:\penalty0 48--54, 2016.
		\newblock \doi{10.1016/j.optlaseng.2015.12.001}.
		
		\bibitem[Kell \textit{et~al.}(2012)Kell, Tyrer, Higginson, Jones, and
		Noden]{Kell_2012}
		Kell, J., Tyrer, J.~R., Higginson, R.~L., Jones, J.~C., and Noden, S.
		\newblock Laser weld pool management through diffractive holographic optics.
		\newblock \emph{Materials Science and Technology}, 28\penalty0 (3):\penalty0
		354--363, 2012.
		\newblock \doi{10.1179/1743284711y.0000000050}.
		
		\bibitem[Funck \textit{et~al.}(2014)Funck, Nett, and Ostendorf]{Funck_2014}
		Funck, K., Nett, R., and Ostendorf, A.
		\newblock Tailored beam shaping for laser spot joining of highly conductive
		thin foils.
		\newblock \emph{Physics Procedia}, 56:\penalty0 750--758, 2014.
		\newblock \doi{10.1016/j.phpro.2014.08.082}.
		
		\bibitem[Han and Liou(2004)]{Han_2004}
		Han, L. and Liou, F.~W.
		\newblock Numerical investigation of the influence of laser beam mode on melt
		pool.
		\newblock \emph{International Journal of Heat and Mass Transfer}, 47\penalty0
		(19-20):\penalty0 4385--4402, 2004.
		\newblock \doi{10.1016/j.ijheatmasstransfer.2004.04.036}.
		
		\bibitem[Safdar \textit{et~al.}(2007)Safdar, Li, and Sheikh]{Safdar_2007}
		Safdar, S., Li, L., and Sheikh, M.~A.
		\newblock Numerical analysis of the effects of non-conventional laser beam
		geometries during laser melting of metallic materials.
		\newblock \emph{Journal of Physics D: Applied Physics}, 40\penalty0
		(2):\penalty0 593--603, 2007.
		\newblock \doi{10.1088/0022-3727/40/2/039}.
		
		\bibitem[Rasch \textit{et~al.}(2019)Rasch, Roider, Kohl, Strau{\ss}, Maurer,
		Nagulin, and Schmidt]{Rasch_2019}
		Rasch, M., Roider, C., Kohl, S., Strau{\ss}, J., Maurer, N., Nagulin, K.~Y.,
		and Schmidt, M.
		\newblock Shaped laser beam profiles for heat conduction welding of
		aluminium-copper alloys.
		\newblock \emph{Optics and Lasers in Engineering}, 115:\penalty0 179--189,
		2019.
		\newblock \doi{10.1016/j.optlaseng.2018.11.025}.
		
		\bibitem[Abadi \textit{et~al.}(2021)Abadi, Mi, Sikström, Ancona, and
		Choquet]{Abadi_2021b}
		Abadi, S. M. A. N.~R., Mi, Y., Sikström, F., Ancona, A., and Choquet, I.
		\newblock Effect of shaped laser beam profiles on melt flow dynamics in
		conduction mode welding.
		\newblock \emph{International Journal of Thermal Sciences}, 166:\penalty0
		106957, 2021.
		\newblock \doi{10.1016/j.ijthermalsci.2021.106957}.
		
		\bibitem[{ANS}()]{Ansys_192}
		\emph{Release 19.2}.
		\newblock {ANSYS Fluent}.
		\newblock URL \url{https://www.ansys.com/}.
		
		\bibitem[Hirt and Nichols(1981)]{Hirt_1981}
		Hirt, C.~W. and Nichols, B.~D.
		\newblock Volume of fluid ({VOF}) method for the dynamics of free boundaries.
		\newblock \emph{Journal of Computational Physics}, 39\penalty0 (1):\penalty0
		201--225, 1981.
		\newblock \doi{10.1016/0021-9991(81)90145-5}.
		
		\bibitem[Ebrahimi \textit{et~al.}(2019{\natexlab{a}})Ebrahimi, Kleijn, and
		Richardson]{Ebrahimi_2019_conf}
		Ebrahimi, A., Kleijn, C.~R., and Richardson, I.~M.
		\newblock The influence of surface deformation on thermocapillary flow
		instabilities in low {Prandtl} melting pools with surfactants.
		\newblock In \emph{{Proceedings of the 5th World Congress on Mechanical,
				Chemical, and Material Engineering}}. Avestia Publishing, 2019{\natexlab{a}}.
		\newblock \doi{10.11159/htff19.201}.
		
		\bibitem[Ebrahimi(2022)]{Ebrahimi_2022t}
		Ebrahimi, A.
		\newblock \emph{Molten Metal Oscillatory Behaviour in Advanced Fusion-based
			Manufacturing Processes}.
		\newblock {PhD} dissertation, Delft University of Technology, Delft, The
		Netherlands, 2022.
		
		\bibitem[Ebrahimi \textit{et~al.}(2019{\natexlab{b}})Ebrahimi, Kleijn, and
		Richardson]{Ebrahimi_2019}
		Ebrahimi, A., Kleijn, C.~R., and Richardson, I.~M.
		\newblock Sensitivity of numerical predictions to the permeability coefficient
		in simulations of melting and solidification using the enthalpy-porosity
		method.
		\newblock \emph{Energies}, 12\penalty0 (22):\penalty0 4360, 2019{\natexlab{b}}.
		\newblock \doi{10.3390/en12224360}.
		
		\bibitem[Ubbink(1997)]{Ubbink_1997}
		Ubbink, O.
		\newblock \emph{Numerical Prediction of Two Fluid Systems with Sharp
			Interfaces}.
		\newblock {PhD} dissertation, Imperial College London (University of London),
		London, United Kingdom, 1997.
		\newblock URL \url{http://hdl.handle.net/10044/1/8604}.
		
		\bibitem[Issa(1986)]{Issa_1986}
		Issa, R.~I.
		\newblock Solution of the implicitly discretised fluid flow equations by
		operator-splitting.
		\newblock \emph{Journal of Computational Physics}, 62\penalty0 (1):\penalty0
		40--65, 1986.
		\newblock \doi{10.1016/0021-9991(86)90099-9}.
		
		\bibitem[Patankar(1980)]{Patankar_1980}
		Patankar, S.~V.
		\newblock \emph{Numerical Heat Transfer and Fluid Flow}.
		\newblock Taylor \& Francis Inc, 1\textsuperscript{st} edition, 1980.
		\newblock ISBN 0891165223.
		
		\bibitem[Hunt(1984)]{Hunt_1984}
		Hunt, J.~D.
		\newblock Steady state columnar and equiaxed growth of dendrites and eutectic.
		\newblock \emph{Materials Science and Engineering}, 65\penalty0 (1):\penalty0
		75--83, 1984.
		\newblock \doi{10.1016/0025-5416(84)90201-5}.
		
		\bibitem[Knapp \textit{et~al.}(2019)Knapp, Raghavan, Plotkowski, and
		DebRoy]{Knapp_2019}
		Knapp, G.~L., Raghavan, N., Plotkowski, A., and DebRoy, T.
		\newblock Experiments and simulations on solidification microstructure for
		{Inconel} 718 in powder bed fusion electron beam additive manufacturing.
		\newblock \emph{Additive Manufacturing}, 25:\penalty0 511--521, 2019.
		\newblock \doi{10.1016/j.addma.2018.12.001}.
		
		\bibitem[Gäumann \textit{et~al.}(2001)Gäumann, Bezen{\c{c}}on, Canalis, and
		Kurz]{Gaeumann_2001}
		Gäumann, M., Bezen{\c{c}}on, C., Canalis, P., and Kurz, W.
		\newblock Single-crystal laser deposition of superalloys:
		processing--microstructure maps.
		\newblock \emph{Acta Materialia}, 49\penalty0 (6):\penalty0 1051--1062, 2001.
		\newblock \doi{10.1016/s1359-6454(00)00367-0}.
		
		\bibitem[Tan and Shin(2015)]{Tan_2015}
		Tan, W. and Shin, Y.~C.
		\newblock Multi-scale modeling of solidification and microstructure development
		in laser keyhole welding process for austenitic stainless steel.
		\newblock \emph{Computational Materials Science}, 98:\penalty0 446--458, 2015.
		\newblock \doi{10.1016/j.commatsci.2014.10.063}.
		
		\bibitem[Mills \textit{et~al.}(1998)Mills, Keene, Brooks, and
		Shirali]{Mills_1998}
		Mills, K.~C., Keene, B.~J., Brooks, R.~F., and Shirali, A.
		\newblock Marangoni effects in welding.
		\newblock \emph{Philosophical Transactions of the Royal Society A:
			Mathematical, Physical and Engineering Sciences}, 356\penalty0
		(1739):\penalty0 911--925, 1998.
		\newblock \doi{10.1098/rsta.1998.0196}.
		
		\bibitem[Steen and Mazumder(2010)]{Steen_2010}
		Steen, W.~M. and Mazumder, J.
		\newblock \emph{Laser Material Processing}.
		\newblock Springer London, 2010.
		\newblock \doi{10.1007/978-1-84996-062-5}.
		
		\bibitem[Assuncao \textit{et~al.}(2012)Assuncao, Williams, and
		Yapp]{Assuncao_2012}
		Assuncao, E., Williams, S., and Yapp, D.
		\newblock Interaction time and beam diameter effects on the conduction mode
		limit.
		\newblock \emph{Optics and Lasers in Engineering}, 50\penalty0 (6):\penalty0
		823--828, 2012.
		\newblock \doi{10.1016/j.optlaseng.2012.02.001}.
		
		\bibitem[Rappaz(1989)]{Rappaz_1989}
		Rappaz, M.
		\newblock Modelling of microstructure formation in solidification processes.
		\newblock \emph{International Materials Reviews}, 34\penalty0 (1):\penalty0
		93--124, 1989.
		\newblock \doi{10.1179/imr.1989.34.1.93}.
		
		\bibitem[Liao and Yu(2007)]{Liao_2007}
		Liao, Y.-C. and Yu, M.-H.
		\newblock Effects of laser beam energy and incident angle on the pulse laser
		welding of stainless steel thin sheet.
		\newblock \emph{Journal of Materials Processing Technology}, 190\penalty0
		(1):\penalty0 102--108, 2007.
		\newblock ISSN 0924-0136.
		\newblock \doi{https://doi.org/10.1016/j.jmatprotec.2007.03.102}.
		
		\bibitem[Kidess \textit{et~al.}(2016)Kidess, Kenjere{\v{s}}, and
		Kleijn]{Kidess_2016_PhysFlu}
		Kidess, A., Kenjere{\v{s}}, S., and Kleijn, C.~R.
		\newblock The influence of surfactants on thermocapillary flow instabilities in
		low {Prandtl} melting pools.
		\newblock \emph{Physics of Fluids}, 28\penalty0 (6):\penalty0 062106, 2016.
		\newblock \doi{10.1063/1.4953797}.
		
		\bibitem[Kurz \textit{et~al.}(2001)Kurz, Bezen{\c{c}}on, and
		Gäumann]{Kurz_2001}
		Kurz, W., Bezen{\c{c}}on, C., and Gäumann, M.
		\newblock Columnar to equiaxed transition in solidification processing.
		\newblock \emph{Science and Technology of Advanced Materials}, 2\penalty0
		(1):\penalty0 185--191, 2001.
		\newblock \doi{10.1016/s1468-6996(01)00047-x}.
		
		\bibitem[He \textit{et~al.}(2021)He, Zhong, Jones, Beuth, and Webler]{He_2021}
		He, Y., Zhong, M., Jones, N., Beuth, J., and Webler, B.
		\newblock The columnar-to-equiaxed transition in melt pools during laser powder
		bed fusion of {M2} steel.
		\newblock \emph{Metallurgical and Materials Transactions A}, 52\penalty0
		(9):\penalty0 4206--4221, 2021.
		\newblock \doi{10.1007/s11661-021-06380-9}.
		
	\end{thebibliography}

\end{document}